# Parallel Implementations Assessment of a Spatial-Spectral Classifier for Hyperspectral Clinical Applications

RAQUEL LAZCANO[1], DANIEL MADROÑAL[1], GIORDANA FLORIMBI[2], JAIME SANCHO[1], SERGIO SANCHEZ[1], RAQUEL LEON[3], HIMAR FABELO[3], SAMUEL ORTEGA[3], EMANUELE TORTI[2], (Member, IEEE), RUBEN SALVADOR[1], (Member, IEEE), MARGARITA MARRERO-MARTIN[3], FRANCESCO LEPORATI[2], (Member, IEEE), EDUARDO JUAREZ[1], (Member, IEEE), GUSTAVO M. CALLICO[3], (Member, IEEE), AND CESAR SANZ[1], (Senior Member, IEEE)

[1]Centre of Software Technologies and Multimedia Systems (CITSEM), Universidad Politécnica de Madrid (UPM), 28031 Madrid, Spain
[2]Department of Electrical, Computer and Biomedical Engineering, University of Pavia, 27100 Pavia, Italy
[3]Institute for Applied Microelectronics (IUMA), University of Las Palmas de Gran Canaria (ULPGC), 35017 Las Palmas de Gran Canaria, Spain

Corresponding author: Raquel Lazcano (raquel.lazcano@upm.es)

This work was supported in part by the Spanish Government through the PLATINO project under Grant TEC2017-86722-C4-4-R, in part by the Regional Government of Madrid through NEMESIS-3D-CM Project under Grant Y2018/BIO-4826, in part by the European Commission through the FP7 Future and Emerging Technologies (FET) Open Program under Grant ICT-2011.9.2, in part by the European Project HELICoiD under Grant 618080, in part by the Canary Islands Government through the Canarian Agency for Research, Innovation and the Information Society (ACIISI) through the ITHaCA project Hyperespectral Identification of Brain Tumors under Grant Agreement ProID2017010164, and in part by the Universidad Politécnica de Madrid through the Programa Propio under Contract RR01/2015 and Contract RR01/2016. The work of S. Ortega was supported in part by the Agencia Canaria de Investigacion, Innovacion y Sociedad de la Información (ACIISI) of the Conserjería de Economía, Industria, Comercio y Conocimiento of the Gobierno de Canarias through the European Social Fund (FSE) [POC 2014-2020, Eje 3 Tema Prioritario 74 (85%)].

**ABSTRACT** Hyperspectral (HS) imaging presents itself as a non-contact, non-ionizing and non-invasive technique, proven to be suitable for medical diagnosis. However, the volume of information contained in these images makes difficult providing the surgeon with information about the boundaries in real-time. To that end, High-Performance-Computing (HPC) platforms become necessary. This paper presents a comparison between the performances provided by five different HPC platforms while processing a spatial-spectral approach to classify HS images, assessing their main benefits and drawbacks. To provide a complete study, two different medical applications, with two different requirements, have been analyzed. The first application consists of HS images taken from neurosurgical operations; the second one presents HS images taken from dermatological interventions. While the main constraint for neurosurgical applications is the processing time, in other environments, as the dermatological one, other requirements can be considered. In that sense, energy efficiency is becoming a major challenge, since this kind of applications are usually developed as hand-held devices, thus depending on the battery capacity. These requirements have been considered to choose the target platforms: on the one hand, three of the most powerful Graphic Processing Units (GPUs) available in the market; and, on the other hand, a low-power GPU and a manycore architecture, both specifically thought for being used in battery-dependent environments.

**INDEX TERMS** Hyperspectral imaging, high performance computing, parallel processing, parallel architectures, image processing, biomedical engineering, medical diagnostic imaging, cancer detection, supervised classification, support vector machines, K-nearest neighbors, principal component analysis, graphic processing unit, manycore.

## I. INTRODUCTION

Hyperspectral Imaging (HSI) collects both spatial and spectral high-resolution information, which is composed of a wide range of wavelengths from across the electromagnetic spectrum. HSI aims to identify and estimate the distribution of materials within a captured scene based on their spectral signatures, which represent the normalized measured surface reflectance at each spectral band [1]. Although

The associate editor coordinating the review of this manuscript and approving it for publication was Fan Zhang.

this technology was originally aimed at remote sensing applications [2], nowadays it has spread to several research fields like forensics [3], food inspection [4], recycling [5], or medicine [6]–[8], among many others. Regarding medical research, the ability to identify the materials within a captured image has been applied in three different environments: *in-vitro*, *ex-vivo* and *in-vivo* studies –i.e., whether the scene is obtained from a resected sample (*in-vitro*, *ex-vivo*) [9]–[11] or directly acquired from the patient (*in-vivo*) [12]–[15]. Specifically, there is a growing research interest related to performing *in-vivo* HSI processing during surgeries to help discerning between cancerous and healthy tissues [16]–[19]. In addition, in order to assist surgeons in locating the margins of the tumor during the operation, a real-time analysis of the acquired hyperspectral (HS) images is required, considering as real-time the time needed by an HS sensor to capture a new image, which will depend on the capturing technology employed by the HS camera [20].

Traditionally, the methodologies proposed to classify each pixel have been uniquely based on its spectrum, regardless of the spatial information [21]. These pixel-wise methodologies can be divided into two categories: spectral feature extraction and spectral classification. The former is based on reducing the spectral dimensionality of the datasets by applying transformations such as the Principal Component Analysis (PCA) [22]. The latter aims to generate a classification model to identify the pixels in relation with different elements –commonly known as *classes*– existing within the image and to depict the boundaries among these classes in the feature space, hence assigning a class to each pixel of the image [23]. In that sense, Support Vector Machines (SVMs) provide robust classification performance when the number of training samples is limited [23], which is common in medical applications [24]. However, it has been recently proven that the combination of both spectral and spatial information for HSI processing can considerably improve the classification results [18], [25]. Specifically, one possible approach to deal with the so-called spatial-spectral (SS) classification aims at refining the pixel-wise classification results by using a K-Nearest Neighbors (KNN) filtering process, which uses both the pixel value and the spatial coordinates. To do so, a one-band representation of the HS image is required, in addition to the probability maps provided by the pixel-wise classifier. Hence, this SS approach is composed of three algorithms: the pixel-wise classifier, an algorithm for obtaining the one-band representation of the image, and the KNN filtering algorithm.

Currently, minimizing the processing time is a crucial task to help medical doctors discriminate between cancerous and healthy tissues. To reach real-time performance, the intrinsic parallelism of this approach must be exploited. For instance, the first two algorithms (PCA and SVM) shall finish before KNN starts its execution. As these algorithms are independent, they can be processed in parallel; in this way, KNN can be executed as soon as possible. In addition to exploiting the intrinsic parallelism of the proposed approach to meet the real-time objective, the time required to process an HS image needs to be minimized, which, in turn, requires an extensive usage of computational resources. Right now, only High-Performance Computing (HPC) architectures are able to provide enough computational power. Two main criteria will be used in this paper to evaluate HPC platforms: processing time and energy consumption. Even though the reduction of the processing time compared to the sequential counterpart –i.e., the speedup– has traditionally been the metric to assess the performance of HPC platforms, energy consumption has also gained importance lately depending on the target application. Although nowadays medical applications do not work under energy constraints, it is easy to foresee future clinical applications where portable real-time HSI processing becomes a tool to support clinical decisions [26], [27]. Therefore, energy-efficient solutions would be advisable.

This paper is focused on the evaluation of the performance and power consumption for two different HS medical applications –with different constraints– implemented on a low-power manycore-based platform and low-power, consumer and scientific-computation Graphic Processing Unit (GPU)-based platforms. The first application is the use of HSI for the intraoperative identification and delineation of *in-vivo* brain tumors, assisting neurosurgeons in the tumor resection during the surgical procedures (from now on called *neurosurgical use case*). In this application, the main goal is to achieve the classification results in the lowest computational time possible, providing near real-time results, without having power limitations within the operating theater. The second application aims to employ HSI for the early detection of skin cancer by using an HS hand-held acquisition and processing system (from now on called *dermatological use case*). In this second case, the system is not intended to be used during surgical procedures, hence, there is a more relaxed deadline. However, this application requires the processing system to work under energy constraints in order to avoid external power connection, being an independent and portable system. Therefore, an implementation with high-energy efficiency would be advisable for this hand-held device.

The main contributions of this paper can be summarized as:
- The implementation of three different algorithms composing a spatial-spectral HS image classifier over different HPC platforms.
- An assessment of these implementations both in terms of processing time and power consumption.
- The use of standard metrics to compare all the implementations and all the platforms.
- The conclusions extracted from applying this metric to assess the fitness of each platform as a function of power consumption, processing time or both.

The rest of the paper is structured as follows. Section II describes the database, the algorithms and the platforms employed in this comparison. Afterwards, Section III explains the algorithm implementations onto each platform and Section IV presents and discusses the obtained results. Finally, Section V draws the main conclusions of this research.

## II. MATERIALS AND METHODS

As described before, the objective of this paper is to compare the performance and power consumption of two different HSI use cases (neurosurgical and dermatological use cases) when implemented onto different high-performance accelerators. This section provides the information needed to better understand the scope of this paper.

### A. HYPERSPECTRAL IMAGE DATABASE

A set of six *in-vivo* HS images have been employed as an example to test the implementations presented in this paper. Three of the images belong to *in-vivo* brain surface affected by cancer and the other three belong to *in-vivo* human skin lesions (normal moles in this case).

On the one hand, brain cancer images (neurosurgical use case) were obtained at the University Hospital Doctor Negrin of Las Palmas de Gran Canaria (Spain) from three different patients with a confirmed grade IV glioblastoma tumor by histopathology. These images were captured using the HS acquisition system developed in [13], [19], and pre-processed following the pre-processing chain presented in [19]. Four steps compose the pre-processing chain: image calibration, noise filtering, band averaging and pixel normalization. The final HS cube is formed of 128 spectral bands, covering the range between 450 and 900 nm [19]. The study protocol and consent procedures were approved by the *Comité Ético de Investigación Clínica-Comité de Ética en la Investigación* (CEIC/CEI) of University Hospital Doctor Negrin and written informed consent was obtained from all subjects. In this use case, the classification goal is to differentiate between tumor tissue and healthy tissue in the exposed brain surface. In addition, two other classes are identified, involving blood vessels and extravassated blood and the other elements, materials or substances presented in the scene (background).

On the other hand, the *in-vivo* HS dermatologic database was obtained with a customized HS acquisition system. Three different lesions from different parts of the body were captured from the same patient and were evaluated by a dermatologist, determining that the lesions were normal moles. The customized HS acquisition system integrates an HS sensor and a monochromatic (MC) sensor, being able to capture an HS image and an MC image with a spatial resolution of 50 × 50 pixels and 1000 × 1000 pixels, respectively. Both the HS and MC images are calibrated and subsequently fused. A final synthetic HS image of 1000 × 1000 pixels and 125 spectral bands is obtained. The synthetic HS image was pre-processed with an image calibration and a noise filtering (removing the extreme bands of the signature), obtaining a final HS cube formed by 100 bands. In this case, the goal is to classify the HS data to provide the dermatologist with a classification map where the different regions in the pigmented lesion are identified as well as the skin area. For the purpose of the work developed in this paper, in these three images only benign pigmented lesions have been included as HS data examples.

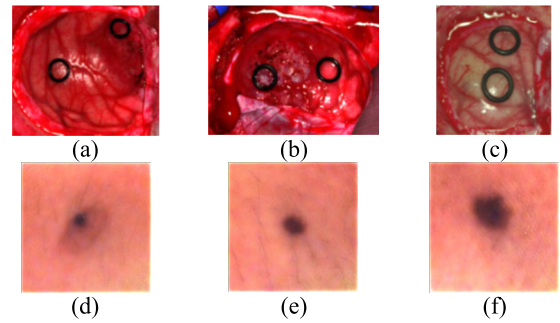

**FIGURE 1.** Synthetic RGB representations of the HS cubes. (a-c) RGB representation of the HS brain image of PB1C1, PB2C1, PB3C1, respectively. (d-f) RGB representation of the HS dermatologic image of PD1C1, PD1C2, PD1C3, respectively.

**TABLE 1.** Specifications of the HS image database.

| Dataset | Image ID | #Pixels | Size (width x height x bands) | Size (MB) |
|---|---|---|---|---|
| Brain | PB1C1 | 219,232 | 496 x 442 x 128 | 107.05 |
|  | PB2C1 | 184,875 | 493 x 375 x 128 | 90.20 |
|  | PB3C1 | 124,033 | 329 x 377 x 128 | 60.56 |
| Derma | PD1C1 | 1,000,000 | 1000 x 1000 x 100 | 381.00 |
|  | PD1C2 | 1,000,000 | 1000 x 1000 x 100 | 381.00 |
|  | PD1C3 | 1,000,000 | 1000 x 1000 x 100 | 381.00 |

Fig. 1 shows the synthetic RGB images of the HS cubes that define the database used in this study, while Table 1 details the characteristics of each HS image. These data can be downloaded in the supplementary material section of the manuscript.

### B. HARDWARE ACCELERATION PLATFORMS

This section provides a brief description of the high-performance accelerators onto which the target applications have been implemented. Due to the large volume of information contained in HS images, there is a clear need of high-performance computing platforms to fulfill real-time requirements. Considering the two use cases under study, it seems obvious that real-time is a strong requirement to help surgeons during tumor resection. However, for the dermatological case, this may not be the main requirement. Instead, it would seem coherent that one of the main requirements for this application could be the energy consumption, as it is not difficult to foresee a future in which this kind of technologies become portable. As a result, different accelerators based on these requirements have been targeted, namely GPUs and manycore platforms. The former ones have been widely used in image processing, usually yielding the most competitive results in terms of speedups. On the other hand, the latter ones provide a solution in which energy consumption is a key factor.

#### 1) HIGH-POWER GPU PLATFORMS

This section presents the characteristics of some of the most powerful GPUs currently available in the market, which gather more than one thousand processing cores and huge memories, in the order of gigabytes. Three NVIDIA desktop GPUs have been chosen in order to cover a wide range of manycore architectures. In particular, authors evaluated one board specifically developed for scientific computations and two consumer GPUs.

The first one is a NVIDIA Tesla K40 GPU, equipped with 2880 cores distributed among 15 Streaming Multiprocessors (SM). The working frequency is 0.875 GHz and the total GDDR5 memory is 12 GB, with a peak bandwidth of 288 GB/s. This GPU is based on the NVIDIA Kepler architecture and it is connected to an Intel i7 3770 CPU (whose working frequency is 3.40 GHz and the RAM memory is 8 GB) through a PCI Express 3.0.

Concerning the consumer GPUs, the tests have been carried out on the NVIDIA GTX 1060 and the NVIDIA RTX 2080 devices. The former is equipped with 1280 cores, working at 1.78 GHz, and organized in 9 SM. It is characterized by 6 GB of GDDR5 memory, with a peak bandwidth of 192 GB/s, and it is based on the Pascal architecture. This board is connected to an Intel i7 8700 CPU (whose working frequency is 3.20 GHz and the RAM memory is 16 GB), through a PCI Express 3.0. The latter is based on the Turing architecture and it includes 2994 cores, organized in 46 SM. The working frequency is 1.71 GHz while the GDDR6 memory is 8 GB, with a peak bandwidth of 448 GB/s. In this case, the GPU is connected to an Intel i9 9900X (whose working frequency is 3.50 GHz and the RAM memory is 128 GB) through a PCI Express 3.0 interface.

#### 2) LOW-POWER GPU PLATFORM
This section details the information related to a new family of GPUs, with smaller computing capabilities, but also with considerably lower power consumption. The selected platform is the NVIDIA Jetson TX2 GPU that is built around the System on Chip (SoC) Tegra X2, which comprises a quad-core ARM Cortex-A57 processor, a dual-core NVIDIA Denver 2 and a NVIDIA Pascal GPU with 256 cores working at a maximum frequency of 1.3 GHz and organized in 2 SM. Unlike discrete GPUs, in this SoC, CPU and GPU share their main 8 GB 128-bit LPDDR4 57.6 GB/s memory, taking advantage of heterogeneity and making possible to avoid heavy memory transfers between CPU and GPU.

#### 3) LOW-POWER MANYCORE PLATFORM
Finally, a manycore architecture, the MPPA-256-N manufactured by Kalray, was employed to compete with GPUs while maintaining a low power consumption. The MPPA-256-N architecture [28] is a single-chip manycore processor that contains 256 processing cores divided in 16 compute clusters whose working frequency can vary between 400 and 600 MHz. In addition, the communications between the clusters and the external memory are managed by 2 quad-core Input/Output (I/O) subsystems. On the other hand, a Network-on-Chip (NoC) is in charge of handling the communication and synchronization between the compute clusters and the I/O subsystems. Regarding the structure of each cluster, they include a Resource Management (RM) core aimed at running the clusters operating system (clusterOS) and managing interrupts and events, a 2 MB local memory and a Direct Memory Access (DMA) to transfer data between the shared memory and the NoC. The MPPA-256-N is connected to a host device with an Intel i7-3820 running at 3.6 GHz.

**TABLE 2.** Specifications of the HS image database.

| Platform | #Cores | Frequency (GHz) | Memory (GB) | Bandwidth (GB/s) | Average Power (W) | CPU |
|---|---|---|---|---|---|---|
| Tesla K40 | 2880 | 0.875 | 12 | 288 | 235 | Intel i7 3770 |
| GTX 1060 | 1280 | 1.78 | 6 | 192 | 120 | Intel i7 8700 |
| RTX 2080 | 2994 | 1.71 | 8 | 448 | 277 | Intel i9 9900X |
| Jetson TX2 | 256 | 1.3 | 8 | 57.6 | 7.5 | ARM Cortex-A57 |
| MPPA-256-N | 256 | 0.55 | 0.032 | 8 | 15 | Intel i7 3820 |

#### 4) PLATFORMS COMPARISON
To conclude this section, Table 2 shows a summary of the main characteristics of each platform presented in this section: three desktop GPUs, a low-power consumption GPU and a manycore platform. As can be observed, the most powerful devices are the three desktop GPUs. In the following sections, the implementation and results for each platform will be presented and compared, trying to find a tradeoff between performance and power consumption.

### C. HS SPATIAL-SPECTRAL (SS) FRAMEWORK
As described in the introduction, the novelty of SS classification techniques is that they improve the spectral-based classification results by incorporating the contextual information into the classifier model; in other words, they combine both the pixel spectral value and its spatial coordinates. Specifically, Fig. 2 presents the block diagram of the SS approach followed in this research work [18], [25]. This approach is composed of two different stages: in the first one, a pixel-wise classification –hereafter, *P*– and a one-band representation –hereafter, *I*– of the image are obtained; after this stage, the KNN filter is applied –hereafter, *O*– to generate a classification map similar to the one shown in Fig. 2 (right), where tumor is identified with red color, healthy tissue with green, hypervascularized tissue with blue, and the other background elements are in black. After this general overview of the SS classification approach, next sections present a brief description of each algorithm included in this method: PCA, SVM and KNN filtering.

#### 1) SVM CLASSIFIER ALGORITHM
The objective of SVM is to classify the contents of a captured scene using the information provided by the spectrum of each pixel [23]. As only the pixel spectrum is considered, this classification can be applied to each pixel independently. In this work, SVM computes the probability of each pixel to belong to each element (or class) of the image. To do that, the algorithm needs two stages: training and classification.

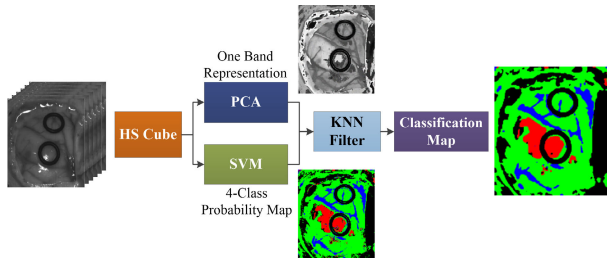

**FIGURE 2.** Block diagram of the SS classification framework.

On the one hand, the training stage is in charge of analyzing a set of samples whose associate classes have been defined a priori by an expert –in this case, a pathologist–. Following a one-vs-one strategy, a set of hyperplanes capable of distinguishing among two of the classes under study is generated. To define each hyperplane –hereafter binary hyperplane–, the criterion applied is to maximize the distance between the classes that it is distinguishing [29].

On the other hand, the classification stage assigns not only a class for each pixel, but also the probability of each pixel to belong to each class. To do so, three steps are needed, and all of them follow a pixel-wise procedure. First, the algorithm computes the distances to the binary hyperplanes and their associated binary probabilities. Secondly, it combines the binary probabilities, following the methodology exposed in [30], which minimizes the implementation complexity. Finally, it selects the class with the highest probability as the one associated to the pixel in order to obtain a preliminary classification result, i.e., without including any spatial filtering. Please, note that this last stage is only performed when using SVM as a standalone algorithm; in this procedure, it feeds the probabilities to the next algorithm. In [31], SVM algorithm is explained in more detail.

### 2) PCA DIMENSIONAL REDUCTION ALGORITHM

As described before, hyperspectral images collect information from across the electromagnetic spectrum, covering a wide range of wavelengths. However, a drawback of this is that this information is deeply correlated between adjacent bands. PCA provides a methodology to minimize this correlation by removing the redundancies among the spectral bands, obtaining the projection that best represents the data in a least-squares sense. To achieve this objective, PCA transforms the original data by means of the eigenvectors of the covariance matrix associated to the original image. As the objective of this algorithm in this application is to obtain a one-band representation of the image, after projecting the image onto the eigenvectors, only the first component is kept.

PCA is divided into four main steps: (i) computation and removal of the average of each band to center the image;

(ii) computation of the covariance matrix of the original image; (iii) extraction of its eigenvectors; and (iv) projection of the original image onto the eigenvectors subspace to keep the first component. In this paper, the method selected for extracting the eigenvectors is the Jacobi approach. It applies successive planar rotations to the largest off-diagonal element to zero it, approximating the original matrix to a diagonal gathering the eigenvalues. These rotations are applied until all the off-diagonal elements are smaller than a provided stop condition. In [32], a description of this method is provided.

### 3) KNN FILTERING ALGORITHM

KNN filter has been extensively used in machine learning applications for HSI [33]. There are several examples in the recent literature [18], [25] that show how to use this filter to improve the classification results provided by classifiers like SVM. To do so, KNN refines the pixel-wise classification probability maps generated by the SVM by matching and averaging non-local neighborhoods. For every pixel in the image, KNN first calculates the K nearest neighbors, and later, it averages the probabilities obtained from the classifier.

To find the $K$ nearest neighbors for a pixel, the distance from that pixel to every other in the image needs to be computed. To calculate them, both the spectral value of the pixel (obtained from the one-band representation of the image) and the spatial coordinates of the pixels are used. A balance parameter, $\lambda$, weights these two data: if $\lambda = 0$, the spatial information is not considered; on the contrary, the greater the value of $\lambda$, the greater the influence given to the local neighborhood. Regarding the number of neighbors, $K$, to be used, it should be highlighted that increasing its value would also increase the computational cost. Consequently, in this work these values have been set to $\lambda = 1$ and $K = 40$, which have been experimentally proven to be a good compromise, as higher values of $\lambda$ and $K$ tend to oversmooth the resulting image [18]. For computing the distance, this research work uses the Euclidean distance, i.e., the squared 2-norm. With this method, the distance from a given pixel with coordinates $(r, c)$ to any other pixel for $\lambda = 1$ is defined as in (1).

$$d(I(rc), I(ij)) = (I_{rc} - I_{ij})^2 + (r - i)^2 + (r - j)^2 \quad (1)$$

where $I_{rc}$ is the normalized pixel value of the image $I$ (the one-band representation of the image) at row $r$ and column $c$, and $I_{ij}$ the value of every other pixel at row $i$ and column $j$.

Once the $K$ nearest neighbors have been found, the filtered result, $O$, is obtained using (2):

$$O(i) = \frac{\sum P(j)}{K}, \quad j \in w_i \quad (2)$$

where $P$ is the original probability map obtained from the SVM classifier and $w_i$ contains the $K$ nearest neighbors of pixel $i$. As a result, there are as many $O$ output probability maps as classes in the SVM classifier. With these optimized probability maps, a final classification map is obtained by assigning the label of the class with the highest probability to each pixel in the image. In [34], KNN filtering algorithm is described in more detail.

## III. PARALLEL IMPLEMENTATIONS

After describing the database, the SS classification approach and the evaluated platforms, this section presents the implementation details of each algorithm onto each platform. It is organized following the same order of the previous section: first, the implementation details of the three desktop GPUs will be presented and, after that, the same will be done for the low power consumption GPU and the manycore. To ease the understanding of this section, Fig. 3 provides a dataflow diagram of the implementations in the GPU-based platforms and the manycore-based platforms.

### A. HIGH-POWER GPU PLATFORMS

The SS classification algorithm has been implemented exploiting three NVIDIA GPUs. These devices have a high number of cores (more than one thousand) and a memory of the order of gigabytes. The limitation of this architecture is the power consumption, which makes this device suitable for applications where this constraint is not relevant, as in the neurosurgical use case. As stated before, Fig. 3.a provides a diagram of the implementation for the three GPUs.

#### 1) SVM ACCELERATION

As previously stated, the SVM classifier independently assigns a class to each pixel. This aspect eases the development of the parallel version of this algorithm which performs a pixel-wise classification. The SVM algorithm consists of three steps: distance estimation, and binary and multiclass probabilities computation, all of them performed on the device.

Before starting the computation on the GPU, the input image and all the variables obtained by the SVM training (which is offline and is not presented in this work) are transferred from the host to the device global memory. After that, the classification starts computing the distances between the samples (pixels) and the hyperplanes, already introduced. This first step is performed exploiting the optimized *cublasSgemm* routine, which belongs to the cuBLAS library [35]. This function computes a matrix-matrix product: in this case, the two inputs are the original image and the matrix containing the hyperplanes data. The distances are stored in a matrix whose size is *number of pixels × number of hyperplanes*. These data are the inputs of the binary and multiclass probabilities computations. They are both computed inside a kernel, whose *grid* dimension is the ratio between the number of pixels and the number of threads present in one *block* of the grid. In this case, this number is set to 32 according to the definition of *warp* provided by NVIDIA. If the remainder of the integer ratio is not zero, the dimension is incremented by one.

As previously explained, the probabilities computation can be pixel-wise parallelized, so the kernel simultaneously computes the binary and the multiclass probabilities for each pixel. First, the kernel faces a binary problem assigning to each pixel the probability to belong to one of the two classes under study. To perform this task, two steps are executed: first, it computes the probability of the sample, associated to one of the classes, through a sigmoid function to the relative distance; secondly, it estimates the probability of the sample to belong to the other class of the classifier. These first stages are executed a number of times equal to the number of hyperplanes. Then the kernel proceeds to compute the multiclass probabilities. In particular, the new probability values are iteratively refined until the difference with the previous iteration is under a threshold or if a maximum error is reached. When one of these two cases is verified, the multiclass probabilities of the sample are computed.

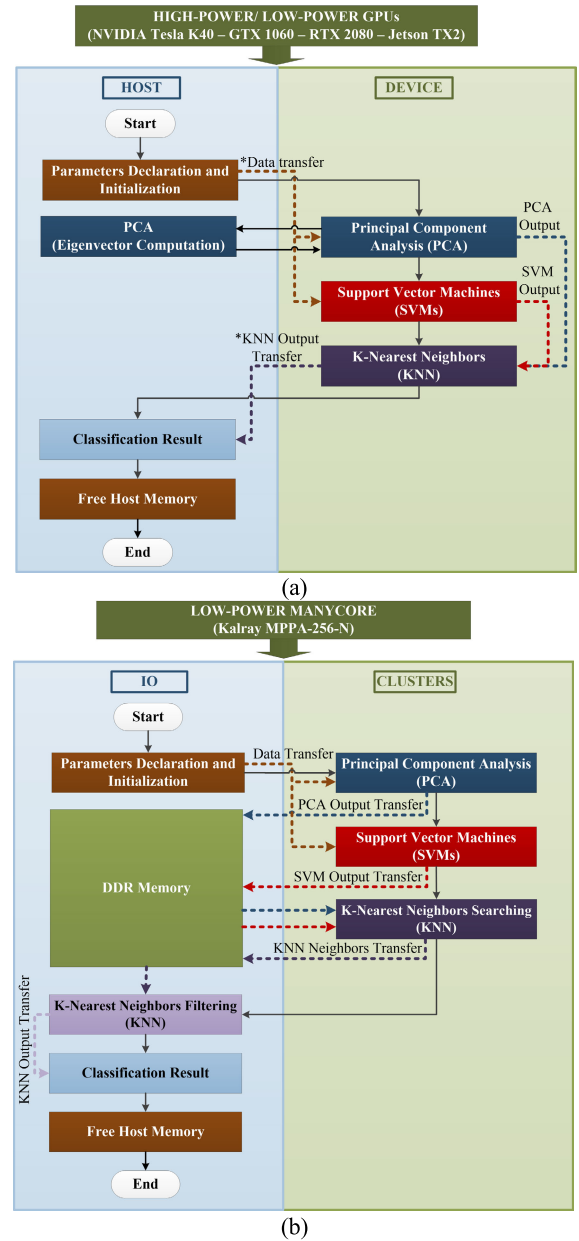

**FIGURE 3.** Dataflow of the (a) GPU-based platform and (b) manycore-based platform parallel implementations. (*) Data transfers are not performed in the Jetson TX2 implementation.

## 2) PCA ACCELERATION

The PCA implementation is divided in four stages: data normalization, covariance matrix computation, Jacobi algorithm execution and projection calculation.

### a: NORMALIZATION

The input of this phase is the image that has been already transferred and stored in the global memory, before performing the SVM algorithm. In this step, two main operations are performed: the averages computation of each band and their subtraction from each element of the corresponding band. Both these steps are performed exploiting the CUDA *streams* [36], that are sequences of operations executed on the device, in the same order they have been issued by the host. Within a stream, the operations are executed in the prescribed order; on the other hand, different streams can overlap their sequences of operations in order to save computational time. In this case, a number of streams equal to the number of bands is created. First, each stream manages the average computation in each row (i.e., each band) of the input matrix. This task exploits the highly optimized *cublasDasum* function that belongs to the cuBLAS library [35]. After, all the averages are computed and subtracted from each element of the matrix row. In this case, each stream executes a custom kernel, whose *grid* dimension is the same as that presented for the SVM algorithm. This kernel removes the average from each element of the corresponding band. Once this step is finished, the streams are removed.

### b: COVARIANCE

The product between the output of the previous step and its transpose matrix, divided by the number of pixels, is the *covariance* matrix. This step exploits another cuBLAS function, *cublasDgemm*, which allows to compute the product between two matrices. It can be noticed that it is not necessary to perform the transpose computation, since it is intrinsically done within the cuBLAS function. The covariance matrix is also stored in the GPU global memory.

### c: JACOBI

As said before, Jacobi is the method chosen to compute the eigenvectors. Authors developed two possible implementations of this step. The former employs a suitable routine of the cuSOLVER library [37] (developed by NVIDIA), called *syevj*. The use of this function requires the declaration and initialization of some variables and arrays. These variables, together with the covariance matrix, represent the inputs of this function, which returns the sorted eigenvectors matrix. The performance analysis of this part of the code shows that the serial eigenvectors computation is faster than the parallel one. For this reason, the authors decided to move this step on the host, executing the serial procedure for the Jacobi method. To perform this step on the host, the covariance matrix is transferred to the CPU; at the end of the computation, only the eigenvectors referred to the principal components P are transferred to the GPU global memory. Despite the increased number of data transfers, the eigenvectors computation on the host is the most efficient and, for this reason, it is the one included in the final system.

### d: PROJECTION

The projection of the input image into the subspace described by the eigenvectors is performed as the product between the image and the eigenvectors of the principal components. This step is computed exploiting another routine of the cuBLAS library. In this case, in fact since the number of the selected principal components is one, this step consists in a product between a matrix and an array performed by the *cublasDgemv* function. If the number of principal components to select is higher than one, the projection is the result of a matrix-matrix product and it is performed by a *cublasDgemm* routine. PCA output is stored in the device global memory.

## 3) KNN ACCELERATION

The basic idea followed in the development of the KNN parallel version is that each CUDA core has to assign a label to each pixel simultaneously. To do this, two main steps have to be performed: the nearest neighbors searching and the filtering phase. As far as the first stage is concerned, the neighbors' searching is done within a restricted area surrounding each pixel, without considering the entire image. Indeed, [18], [34] use a window characterized by 14 rows of pixels, instead of the entire image, accelerating the computation without modifying the classification accuracy. For this reason, the first stage performed by the KNN parallel version is the definition of the pixels windows.

### a: NEIGHBOR SCAN

A custom kernel computes the borders and the size of the windows in parallel through the pixels. Notice that it is important to know these parameters before starting the KNN computation because they allow to evaluate the neighbors for each pixel simultaneously. The nearest neighbors searching is performed by a custom kernel, which is characterized by a number of threads equal to the number of pixels. This kernel exploits the PCA output, already stored in the global memory, in the distances computation. For each thread of the kernel, two arrays (whose dimension is equal to the number of neighbors) are stored in the local memory: the former contains the distances, each one initialized with a huge value; the latter contains the neighbors' indexes. Every thread computes the Euclidean distances between the assigned pixel and the ones within its window. After each computation, if the distance is smaller or equal to the last element of the array, it will be stored in this last position. Once a new distance is stored, the algorithm sorts the elements of the array in an ascending order, keeping track of their indexes. The indexes of the selected neighbors are the output of the kernel and they will be copied to the global memory. In particular, a matrix containing all the neighbors of each pixel is allocated in the global memory and each thread can store its neighbors at the end of the kernel. Once all the neighbors are stored, a new kernel can perform the filtering phase considering the SVM output, already in the global memory.

#### b: FILTERING
First, each thread in the kernel copies the SVM probabilities of their corresponding neighbors from the global to the local memory. For each class, each thread computes the sum of the SVM probabilities of all the neighbors of the reference pixel. The label corresponding to the class with the highest sum of probabilities is assigned to the pixel and stored in an array in the global memory. The final output is an array containing all the pixel labels.

### B. LOW-POWER GPU PLATFORMS
This section includes the implementation details for the three algorithms using the embedded platform NVIDIA Jetson TX2. Although this system also uses a GPU to parallelize the heaviest computational processes, the computational power and the amount of memory is not as high as in the NVIDA desktop GPUs, being necessary in some occasions the special tailoring of some functions. In addition, this system benefits from sharing the same memory for host and device, avoiding memory transfers between the embedded CPU and GPU. This fact introduces the possibility of using the CPU for non-parallelizable algorithms without drawbacks and the possibility of running different processes in CPU and GPU concurrently. This aspect is also represented in Fig. 3.a, as the main difference among the GPU-based implementations is that Jetson TX2 does not perform data transfers.

#### 1) SVM ACCELERATION
The SVM implementation is split up into two kernels, where the first one computes the binary probabilities –i.e., first step described in Section II.C– and the second one performs the decoupling of the class probabilities –steps 2 and 3 described in Section II.C.

#### a: BINARY PROBABILITIES
The first kernel calculates the binary probabilities by first computing the distances of the hyperspectral images to all the separating hyperplanes and then transforming them to pairwise probabilities using an adapted sigmoid function [31]. This is implemented as a naive matrix multiplication between the weight matrix and the hyperspectral image. The problem space is mapped in blocks of 64 threads, where each thread computes all the values for all the classifiers. Based on [38], we increased the workload by leveraging the number of pixels each thread processes, reducing the shared memory occupancy. We have obtained the best results when each thread processes eight pixels and the occupancy is set to 25%. The SVM model (except the weights matrix) is stored in constant memory, a type of read-only memory space optimized for uniform access across threads in a warp. The weights matrix and the hyperspectral image are stored in the global memory space but cached in the L1 and L2 caches.

#### b: DECOUPLE
In this step, class probabilities are estimated by combining the pairwise coupled probabilities [31] computed on the previous kernel. Just as in the previous step, a block of 64 threads is used, each one computing the class probabilities for one pixel. All auxiliary variables are stored on registers.

#### 2) PCA ACCELERATION
Since PCA comprises four well-differentiated stages, the implementation is also divided into four main functions: data normalization, covariance matrix computation, Jacobi algorithm execution and projection calculation.

#### a: NORMALIZATION
This functionality is implemented using two different kernels: one calculating the average per band and another subtracting each pixel from the average. The first kernel makes use of the GPU shared memory to perform an average over the different bands using a number of blocks equal to the number of bands and 1024 threads per block. The idea is that every thread within a block performs an addition to the next chunk of band, i.e., the following 1024 values, until the band's end is reached. At this point, each thread stores a portion of the required sum, hence it is needed to combine its values to obtain the average of the complete band. For that purpose, the shared memory is used to perform the so-called average reduction [39]. As the values in shared memory can be accessed by all the threads within a block, it is possible to use only the lower half of threads to sum its values with the higher ones. Repeating this process ($log_2 1024 =$)10 times, the first thread ends up with the sum band value. The second kernel uses a different thread for each different SS pixel in the hyperspectral image in order to subtract the corresponding band sum divided by the spatial size of the image.

#### b: COVARIANCE
The process of creating the covariance matrix consists in the multiplication of each centered band with itself and all the others in order to obtain a symmetric matrix. For this reason, although the process can be summarized as the multiplication of a matrix by its transposed, performing the transposed step is not needed. To benefit from these implications, we used the *syrk* BLAS routine, which performs a rank-n update of an n-by-n symmetric matrix and it is implemented in the cuBLAS library.

#### c: JACOBI
Since this process is iterative and strongly dependent on the previous iterations data, several experiments revealed that the GPU paradigm is not able to accelerate the algorithm with such a reduced number of bands [40]. Therefore, and taking into account that no memory copies are needed between CPU and GPU, this process has been developed in the CPU.

#### d: PROJECTION
In this stage, the centered HS image represented as a matrix is multiplied by the principal eigenvector, resulting in a matrix-vector multiplication. This operation is performed mapping one thread per spatial pixel and iterating across the bands within them. By this way, each thread stores the result of multiplying and accumulating all the spectral bands of a spatial pixel with the main eigenvector.

#### 3) KNN ACCELERATION

The implementation of the KNN filtering algorithm is divided in two stages: neighboring scanning process and mean computation.

##### a: NEIGHBOR SCAN

The first step calculates the distances from one pixel to the rest in the rectangular window. Each thread computes all the distances for one origin pixel as stated on Equation (1). As consecutive threads read consecutive pixels and move forward one pixel on each iteration, pixels are cached in L1 cache. The second kernel searches for the K nearest neighbors in the distances space by using a partial heapsort with online setting. The approach per pixel is as follows: initially, the first K distances are read and turned into a max-heap using the *heapify* procedure. Then, the remaining distances are read sequentially and compared against the top parent leaf of the heap. If the read distance is lower than the leaf, the leaf is replaced and the *heapify* procedure is called again to reorder the heap. The upper-bound complexity of the algorithm is $O\left(n \log_2 K\right)$ where n is the number of distances per pixel and K is the number of neighbors. Local memory space is chosen for storing the heap because, although located in global memory, it is heavily cached in L1 and L2 caches. Since the number of distances per pixel is high, the amount of memory needed to store them all might exceed the system memory. Consequently, a batch approach is chosen for the first two kernels with a batch size of 10 rows. A buffer is used to store temporarily the distances and pass them to the second kernel. This buffer is fully filled on each batch iteration by the distance kernel except when the rectangular window is on the top or the bottom of the image; in this case, the buffer overlaps with the start and the end phase and is filled with the maximum value representable by a *float* type for the correct execution of the neighbors scan stage. In our case, the number of rows of the buffer is bigger than the span of the rectangular window so the buffer clearing only takes place at the first and the last batches.

##### b: MEAN COMPUTATION

For the third kernel, each thread is assigned to a pixel that computes the mean of the class probabilities of the SVM for its neighbors and selects the class with the highest probability.

#### C. LOW-POWER MANYCORE PLATFORM

In this section, the different approaches regarding the implementation of each algorithm on the MPPA-256-N are detailed. Generally speaking, the main constraint existing within this platform is the reduced amount of memory available in each cluster. Although the total amount of memory is 2 MB, there is only 1 MB left for data storage, since the rest is reserved for the operation system and the application itself. Consequently, all the implementations have been aimed at optimizing the use of this memory and reducing the total amount of transmissions from the I/O subsystem to the clusters, since the host is only used for transferring the image to the I/O subsystem.

Additionally, it should be highlighted that, compared to previous implementations on the platform [31], [32], this paper introduces the usage of a new communication library based on unilateral and asynchronous data transmission. This new library allows the developer to simplify the communication procedure while maintaining the performance [41]. As a summary, a dataflow diagram of the implementation is provided in Fig. 3.b.

#### 1) SVM ACCELERATION

Concerning SVM acceleration using the MPPA-256-N, the structure of the algorithm in terms of data parallelism is considered pixel-wise. That is, each pixel of the image can be processed in parallel. Hence, the implementation of this algorithm has been planned to optimize both data transmissions and clusters memory usage.

To do so, and thanks to the usage of the asynchronous communication library, each processing element inside each cluster *reads* data from the I/O subsystem memory, processes its corresponding block of 64 pixels and *writes* the result on the I/O subsystem memory again. With this structure, both communication and processing can be parallelized and, additionally, a 16-buffer technique is automatically employed to communicate data as the asynchronous library receives the *reading* request and performs the communication as soon as the DMAs finish the previous transmission.

#### 2) PCA ACCELERATION

The PCA algorithm, instead, depends on the step of the algorithm that should be accelerated. To this regard, different strategies have been considered depending on the parallelism within each step of the algorithm.

##### a: NORMALIZATION

In the case of the normalization step, a fixed size of 64 KB is considered as the standard block size when transmitting data from the I/O to the compute clusters. Specifically, 16 K pixels of the same band are grouped in each block, so, when 16 processing elements are working in parallel, 1 MB of data is used to compute the average of one band and to remove it from each pixel of the band. Likewise, each cluster is in charge of performing the normalization of one band and, to do so, each processing element is *reading* one block of pixels from the I/O subsystem memory and accumulating the result. After all the blocks composing one band are processed, the average of the band is computed (i.e., a reduction strategy is considered in this stage). Once the average of the band is obtained, the processing elements *read* the pixel blocks, the average is subtracted from them and the result is *written* on the I/O subsystem in both the same order of the reading and performing a transposed of the image that will be used during the *Projection* step. To perform this transposition, the asynchronous communication library allows the developer to perform spaced *readings/writings* so the transposition of a matrix is carried out together with the communication, saving processing time.

*b: COVARIANCE*

Secondly, during the covariance computation step, the implementation takes advantage of the nature of the result of multiplying a matrix by its transposed, which is a symmetric matrix. In this situation, the total amount of vector multiplications required is almost halved. In this part of the algorithm, each cluster is in charge of computing the vector multiplications associated to each row of the result and iterates until fulfilling the covariance matrix. First, each row/column/band of the image is divided in blocks of the same size as in the previous stage (64 KB). After that, the block associated to the row is *read* from the I/O subsystem by one of the processing elements of each cluster. Likewise, each processing element iterated *reading* column blocks until all of them are processed. Then, the following row is *read* and all the processing elements repeat the same procedure until all the row corresponding to the result is obtained. Finally, the result is written in both a row- and a column-wise order on the resulting covariance matrix, using both the normal and the spaced *writings* modes of the asynchronous communication library.

*c: JACOBI*

Thirdly, due to the strong data dependency existing within the Jacobi algorithm, it has been decided to compute it independently in each cluster. As this algorithm aims at diagonalizing the covariance matrix, it only required 128 × 128 values when analyzing brain cancer images and 100 × 100 when using dermatology images, totaling for 64 KB and 39 KB of data, respectively. Inside each cluster, the processing element 0 is in charge of selecting the elements to be zeroed, and afterwards all the processing elements within the cluster compute the result after zeroing each element. This process is repeated iteratively until the diagonalization is finished. After that, only the first eigenvector (the one associated to the first principal component) is kept in memory.

*d: PROJECTION*

Finally, the strategy considered for the *Projection* is equivalent to the one for the SVM algorithm. The image is divided in blocks of 64 pixels (64 × 128 float values), totaling for 32 KB. Hence, when using the 16 processing elements inside each compute cluster, a total amount of 512 MB of memory will be used. The main difference of this algorithm with the SVM is the input data, as this algorithm uses the transposed of the normalized matrix as input data, while the SVM algorithm uses the transposed of the original image. That is why both algorithms need to be processed in different time slots and their communication parts are not shared.

*3) KNN ACCELERATION*

The strategy followed to accelerate the KNN algorithm is almost equivalent to the one used for SVM. In this case, the pixels can also be processed in parallel, but the amount of data required to process one pixel is not proportional to the one required to process N pixels. That is, as a band window has been set to perform the searching of the K nearest neighbors, to process one pixel 14 lines of the image are required, while, for example, only 14 lines and 1 pixel are required when processing 2 pixels.

To be compliant with the memory constraints of the platform, a conservative approach of using blocks of 64 pixels have been chosen to implement the KNN searching. With this pattern, the memory usage in the worst case (dermatology images) is the following: 1) each pixel of the image is represented as a set of 12 bytes (PCA output value, row, column and index). Consequently, the input data is 1000 (samples) × 14 (lines composing the searching band) + 64 pixels × 12 bytes, totaling for 164.81 KB; 2) For each pixel, a total amount of 40 neighbors are computed, so 10 KB are reserved for output data in this algorithm.

With this implementation, images with a resolution equivalent to 4 K could be analyzed using this algorithm, as they would use only 672 KB of input data. Specifically, each cluster loads the data necessary to analyze each block of 64 pixels and each processing element looks for the neighbors associated to a subset of each block of pixels, e.g., when using 16 processing elements per cluster, each processing element processes 4 pixels of each block.

## IV. EXPERIMENTAL RESULTS AND DISCUSSION

After describing the implementation details of each algorithm for each platform, this section presents a thorough analysis of the obtained results, focusing on two different aspects: performance –i.e., processing time– and power consumption. However, before presenting the results gathered for each platform and implementation, first some common ground must be set to simplify the understanding of the remaining section. Regarding performance, it is going to be assessed by measuring the execution time per algorithm. Regarding power consumption, this metric is going to be obtained considering the application as a whole. Finally, it must be highlighted that all the measurements presented in this section are the result of an average of 20 repetitions, with a standard deviation smaller than 1% for the execution times.

To simplify the comparison among all the implementations, three different Figures of Merit (FoMs) have been defined. These FoMs link both the global time and the power consumption of the SS classification. *FoM* 1 is a general index, expressed as (3), inversely proportional to the product between time ($T$) and power ($P$). *FoM* 2 and *FoM* 3 are similar to the first index, but the former gives more importance to the time while the latter favors the power consumption. They are expressed as (4) and (5), respectively.

$$FoM\,1 = \frac{1}{T*P} \quad (3)$$

$$FoM2 = \frac{1}{T^2 * P} \quad (4)$$

$$FoM3 = \frac{1}{T * P^2} \quad (5)$$

The choice of these three FoMs is due to the fact that authors want to provide a general evaluation of different technologies, together with the possibility to favor some specific aspects related to an application. For these reasons, $FoM1$ gives equal importance to the processing time and power consumption, presenting a general metric that does not consider specific application constraints. On the contrary, $FoM2$ gives more importance to the processing time, so it is suitable for evaluating applications where time constraints are critical, as is the case of the neurosurgical use case. During a neurosurgery, it is crucial to not increase the duration of the procedure and, for this reason, a real-time classification is mandatory. However, the operating room has not power supply restrictions and, as $FoM3$ favors the power consumption, it can be used to evaluate those applications where this aspect represents a constraint. Considering the dermatological use case, the classification is not performed during a surgery, but it is evaluated by a portable device, powered by a battery. Therefore, the prevalent aspect is the power consumption.

The section is organized as follows: first, the individual results of each implementation and platform are presented in Sections IV.A, B and C. After that, Section IV.D compares all of them and extracts the conclusions.

### A. HIGH-POWER GPU PLATFORMS

This section presents the classification results of the HS image classification, obtained by exploiting three desktop NVIDIA GPUs to accelerate the heaviest parts of the SS classification algorithm. As already discussed in Section III.A, the final version of this system performs the PCA, SVM and KNN algorithms on the device. Only one part of the computation is executed on the host: in fact, the Jacobi method is faster if computed on the CPU, despite the increased number of data transfers.

Table 3 summarizes the performances obtained exploiting the three desktop NVIDIA GPUs used in this work, while Fig. 4.a shows the computational times related to the single algorithms (in logarithmic scale). It must be highlighted that those times refer only to the effective execution of the algorithms, without considering the cuBLAS context and the streams creation, the inputs transfers, the device memory allocation and deallocation. Analyzing these times, it is possible to conclude that the KNN algorithm is the heaviest part from the computational point of view, while the SVM is the fastest one. Moreover, the PCA times are calculated considering also the time of the eigenvectors computation performed on the host (including the data transfers). This time represents a very low percentage of the entire PCA computation: for example, considering one of the biggest images of the database (PD1C1) and the NVIDIA RTX 2080 GPU, the time related to this phase is 1 ms compared to the 23 ms of the entire PCA execution. The column *Time*, in Table 3, shows the computational times of the SS classification system, including the algorithm times (already presented) and the ones related to the cuBLAS context and the streams creation, the inputs transfers, the device memory allocation and deallocation. Analyzing these classification results, it is possible to make a comparison between the three boards. It is worth noting that all the boards provide a real-time classification for all the images. The device with the highest performance is the NVIDIA RTX 2080, which takes less than 2.5 s to evaluate the dermatological images, i.e., the biggest ones. Indeed, this board is characterized by the highest ratio between the number of cores and the working frequency, and it is based on the Turing architecture, which is the most recent among the three considered devices. If the neurological images classification is taken into account, the Tesla K40 shows better performance than the GTX 1060 but, concerning the dermatological images, which have the highest number of pixels, this result is reversed. This aspect may be due to the large number of accesses to the device global memory during the KNN computation: in fact, in the neighbors searching and in the filtering phases the PCA and SVM results, stored in the global memory, have to be read several times, increasing the classification time. The GTX 1060 is based on a more recent architecture and it is characterized by a higher working frequency compared to the Tesla K40, therefore it can perform the memory loading in a more efficient way, reducing also the computational time.

**TABLE 3.** High-power GPU platforms (Tesla K40, GTX 1060, RTX 2080) implementation results.

| Image ID | Device | Time [s] | Av. Power [W] | FoM1 [mJ$^{-1}$] | FoM2 [mJ$^{-1}$/s] | FoM3 [mJ$^{-1}$/W] |
|---|---|---|---|---|---|---|
| PB1C1 | Tesla K40 | 1.77 | 36.58 | 15.36 | 8.63 | 0.420 |
|  | GTX 1060 | 1.99 | 38.79 | 12.92 | 6.47 | 0.333 |
|  | RTX 2080 | 1.23 | 66.09 | 12.21 | 9.85 | 0.185 |
| PB2C1 | Tesla K40 | 1.49 | 34.04 | 19.71 | 13.23 | 0.579 |
|  | GTX 1060 | 1.75 | 38.81 | 14.68 | 8.36 | 0.378 |
|  | RTX 2080 | 1.13 | 65.89 | 13.42 | 11.86 | 0.204 |
| PB3C1 | Tesla K40 | 1.10 | 35.68 | 25.45 | 23.11 | 0.713 |
|  | GTX 1060 | 1.36 | 29.30 | 24.94 | 18.23 | 0.852 |
|  | RTX 2080 | 1.07 | 58.42 | 15.95 | 14.86 | 0.273 |
| PD1C1 | Tesla K40 | 8.67 | 28.95 | 3.98 | 0.45 | 0.138 |
|  | GTX 1060 | 2.90 | 17.80 | 19.34 | 6.65 | 1.087 |
|  | RTX 2080 | 2.49 | 88.45 | 4.54 | 1.82 | 0.051 |
| PD1C2 | Tesla K40 | 8.66 | 28.98 | 3.98 | 0.45 | 0.137 |
|  | GTX 1060 | 2.87 | 29.44 | 11.83 | 4.12 | 0.402 |
|  | RTX 2080 | 2.46 | 88.51 | 4.57 | 1.85 | 0.052 |
| PD1C3 | Tesla K40 | 8.68 | 36.45 | 3.16 | 0.36 | 0.087 |
|  | GTX 1060 | 2.87 | 14.27 | 24.38 | 8.48 | 1.709 |
|  | RTX 2080 | 2.49 | 88.20 | 4.54 | 1.81 | 0.051 |

Table 3 shows the average power consumed by the three devices. It has been measured through the NVIDIA System Management Interface (*nvidia-smi*) tool, which allows sampling the power during the execution on the device. The time step chosen for the sampling is 1 ms. It must be highlighted that this tool provides a measure of the power consumed by the device. As said before, the SVM and the KNN are entirely executed on the GPU but, in the PCA, the eigenvectors computation is performed on the host. For this reason, the *Av. Power* does not take into account this step. In this context, it must be noticed that the eigenvectors computation represents the 0.04% of the entire classification time (in the case of the image PD1C1 and the RTX 2080 GPU), and therefore it is negligible. Moreover, authors tried to measure the power consumption of the SS classification considering the PCA entirely performed on the device (so, computing the eigenvectors with the cuSOLVER function). The values obtained are comparable with the ones presented in Table 3. This further result demonstrates that the power consumption related to this specific step is negligible if compared to the entire execution.

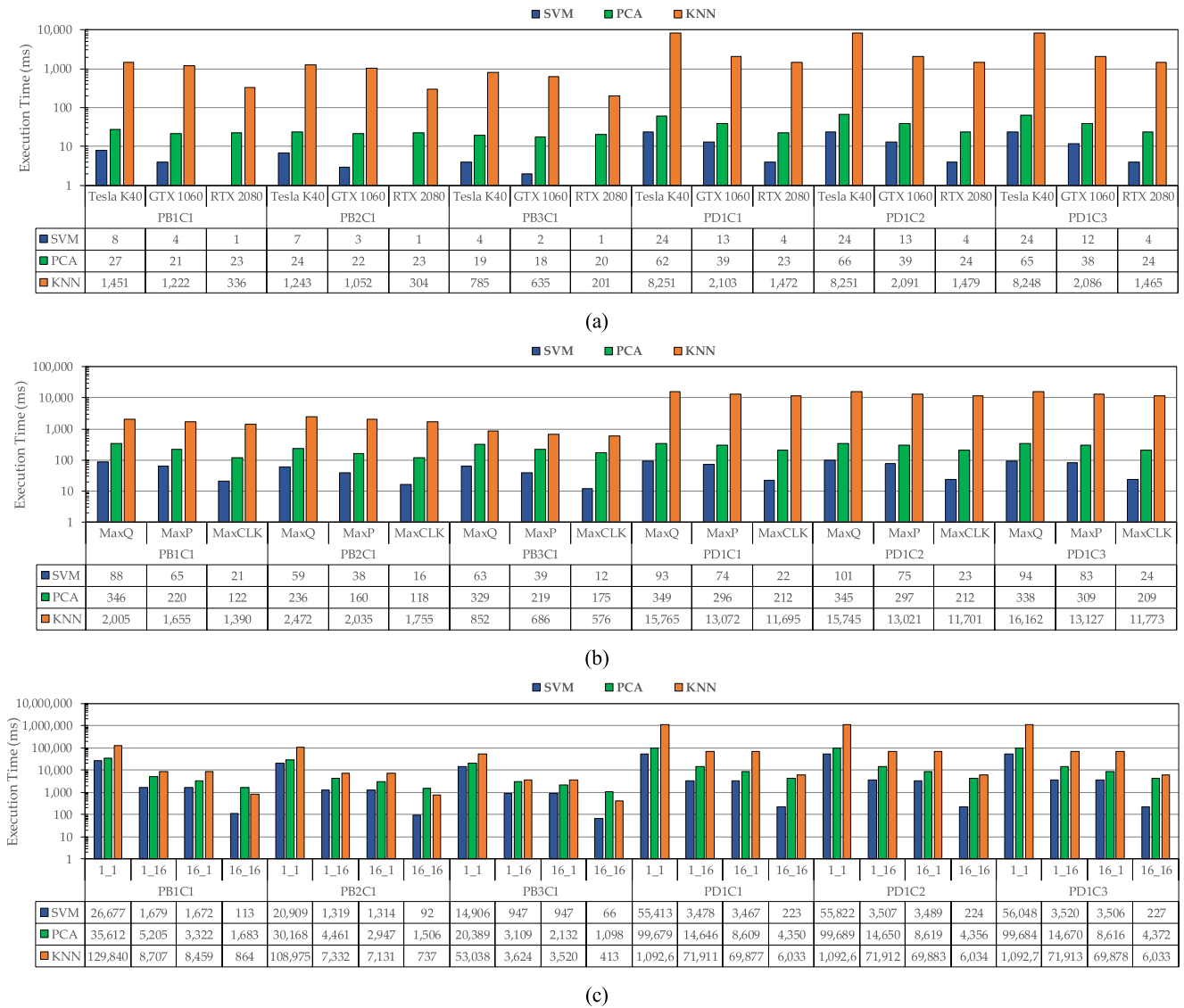

**FIGURE 4.** Execution time of each algorithm per platform implementation for each HS image of the database (representation in logarithmic scale). (a) High-power GPU platforms results. (b) Low-power GPU platform results. (c) Low-power manycore platform results.

Considering the three devices in Table 3, the GTX 1060 is mainly the least power-demanding device. It is based on the Pascal architecture that is optimized for the power consumption. Concerning the dermatological images, it is possible to notice that both GTX 1060 GPU and Tesla K40 GPU are more efficient than RTX 2080 GPU, mainly due to the higher number of cores that works at a higher frequency.

Finally, analyzing the FoMs in Table 3, it is possible to notice that the Tesla K40 GPU is the one with the best performance for neurological images, while GTX 1060 GPU provides the best performance for dermatological images due to the low values of power consumption. *FoM* 1 confirms this consideration. Indeed, for the dermatological images, the GTX 1060 GPU achieves lower processing times than the Tesla K40, as explained before. Moreover, its power consumption values ensure a higher *FoM* 1 than the Tesla K40 one.

It is important to highlight that there is a large difference between *FoM* 2 and *FoM* 3 values. This was expected due to the characteristics that these desktop GPUs presents. In fact, as said before, they are more suitable for those applications where the power supply is not strictly limited and where fast elaborations are needed. Therefore, considering these devices, $FoM2$ better describes the performances of this application. Considering this index, Tesla K40 GPU shows the highest values in the neurological images classification, while the GTX 1060 GPU better performs considering the dermatological images.

**TABLE 4.** Low-power GPU platform (Jetson TX2) implementation results.

| Image ID | Device | Time [s] | Av. Power [W] | FoM1 [mJ$^{-1}$] | FoM2 [mJ$^{-1}$/s] | FoM3 [mJ$^{-1}$/W] |
|---|---|---|---|---|---|---|
| PB1C1 | MaxQ | 3.28 | 2.33 | 130.22 | 39.65 | 55.69 |
| | MaxP | 2.50 | 4.46 | 89.36 | 35.65 | 20.01 |
| | MaxCLK | 2.20 | 4.95 | 91.80 | 41.72 | 18.54 |
| PB2C1 | MaxQ | 3.56 | 2.69 | 104.10 | 29.19 | 38.64 |
| | MaxP | 2.78 | 4.32 | 82.93 | 29.73 | 19.18 |
| | MaxCLK | 2.51 | 5.38 | 73.90 | 29.38 | 13.73 |
| PB3C1 | MaxQ | 1.99 | 1.94 | 258.37 | 129.68 | 132.99 |
| | MaxP | 1.44 | 3.01 | 230.45 | 159.90 | 76.53 |
| | MaxCLK | 1.32 | 4.14 | 181.67 | 136.91 | 43.79 |
| PD1C1 | MaxQ | 18.15 | 3.36 | 16.39 | 0.90 | 4.87 |
| | MaxP | 14.61 | 4.86 | 14.05 | 0.96 | 2.88 |
| | MaxCLK | 13.17 | 6.37 | 11.91 | 0.90 | 1.87 |
| PD1C2 | MaxQ | 18.10 | 3.40 | 16.23 | 0.89 | 4.77 |
| | MaxP | 14.55 | 4.84 | 14.19 | 0.97 | 2.93 |
| | MaxCLK | 13.18 | 6.34 | 11.95 | 0.90 | 1.88 |
| PD1C3 | MaxQ | 18.44 | 3.53 | 15.33 | 0.83 | 4.33 |
| | MaxP | 14.68 | 4.81 | 14.15 | 0.96 | 2.94 |
| | MaxCLK | 13.24 | 6.34 | 11.90 | 0.89 | 1.87 |

**TABLE 5.** Low-power manycore platform (MPPA-256-N) implementation results.

| Image ID | CC-PE | Time [s] | Av. Power [W] | FoM1 [mJ$^{-1}$] | FoM2 [mJ$^{-1}$/s] | FoM3 [mJ$^{-1}$/W] |
|---|---|---|---|---|---|---|
| PB1C1 | 1-1 | 192.68 | 8.47 | 0.61 | 0.003 | 0.072 |
| | 1-16 | 16.14 | 8.27 | 7.49 | 0.464 | 0.906 |
| | 16-1 | 14.02 | 11.16 | 6.39 | 0.456 | 0.573 |
| | 16-16 | 3.23 | 10.82 | 28.61 | 8.859 | 2.644 |
| PB2C1 | 1-1 | 160.52 | 8.34 | 0.74 | 0.005 | 0.090 |
| | 1-16 | 13.58 | 8.27 | 8.90 | 0.656 | 1.077 |
| | 16-1 | 11.87 | 11.20 | 7.52 | 0.634 | 0.672 |
| | 16-16 | 2.82 | 10.62 | 33.39 | 11.841 | 3.144 |
| PB3C1 | 1-1 | 88.65 | 8.37 | 1.34 | 0.015 | 0.161 |
| | 1-16 | 7.99 | 8.18 | 15.30 | 1.915 | 1.870 |
| | 16-1 | 6.93 | 10.85 | 13.30 | 1.919 | 1.226 |
| | 16-16 | 1.91 | 9.90 | 52.88 | 27.688 | 5.342 |
| PD1C1 | 1-1 | 1,249.92 | 8.28 | 0.09 | 0.000 | 0.012 |
| | 1-16 | 92.16 | 8.48 | 1.28 | 0.014 | 0.151 |
| | 16-1 | 84.08 | 11.61 | 1.02 | 0.012 | 0.088 |
| | 16-16 | 12.73 | 12.40 | 6.33 | 0.498 | 0.511 |
| PD1C2 | 1-1 | 1,250.33 | 8.18 | 0.09 | 0.000 | 0.012 |
| | 1-16 | 92.19 | 8.49 | 1.27 | 0.014 | 0.150 |
| | 16-1 | 84.12 | 11.53 | 1.03 | 0.012 | 0.089 |
| | 16-16 | 12.74 | 12.61 | 6.22 | 0.489 | 0.494 |
| PD1C3 | 1-1 | 1,250.59 | 8.04 | 0.09 | 0.000 | 0.012 |
| | 1-16 | 92.23 | 8.49 | 1.27 | 0.014 | 0.150 |
| | 16-1 | 84.12 | 11.45 | 1.03 | 0.012 | 0.091 |
| | 16-16 | 12.76 | 12.61 | 6.21 | 0.487 | 0.493 |

### B. LOW-POWER GPU PLATFORMS

The implementation described in Section III-B has been used to conduct several experiments using the HS images selected in this paper. As in the previous subsection, the evaluation is summarized in Table 4 including the three aforementioned figures of merit, while Fig. 4.b presents the processing time needed stage by stage (in logarithmic scale). Power measurements correspond to the GPU and DRAM consumptions except for the Jacobi method, which replaces GPU with CPU consumption. In addition, three different clock configurations are assessed: (i) *MaxQ* prioritizes energy efficiency by constraining the GPU clock up to 800 MHz and the memory clock up to 1.30 GHz; (ii) *MaxP* leverages between performance and energy efficiency by constraining the GPU clock up to 1.12 GHz and the memory clock up to 1.60 GHz; finally, (iii) *MaxCLK* fixes the GPU clock to 1.30 GHz and the memory clock to 1.87 GHz, improving performance at the expense of decreasing power efficiency.

Results show that, as expected, *MaxCLK* obtains the best performance in terms of computing times followed by *MaxP* and then by *MaxQ*, achieving the opposite in terms of average power. Nevertheless, focusing on $FoM1$, it can be observed that *MaxQ* obtains better results for all the image set, which means that the balance between time and power is optimum when using this mode. Regarding $FoM2$, results vary depending on the assessed image, with similar ratios for the different modes. To end up, $FoM3$ behaves as expected, achieving the best results *MaxQ* mode in every case. Taking into account the results, *MaxQ* mode is considered the best option to seize the platform. On the other hand, if execution time is key concern, *MaxCLK* is the preferred mode.

### C. LOW-POWER MANYCORE PLATFORM

Finally, the results associated to the implementation using the many-core architecture are analyzed. As can be seen in Table 5, four different configurations have been addressed to completely analyze the application behavior on the MPPA-256-N. Specifically, a serial (1-1), a multi-core parallelism (1-16), a multi-cluster parallelism (16-1) and a fully many-core parallelism (16-16) configurations are studied and displayed on the column CC-PE.

When comparing the workload distributions of each algorithm with each configuration –Fig. 4.c (in logarithmic scale)–, the obtained speedups vary considerably. In average, the speedup of the SVM algorithm reaches 231x when compared to the sequential version, while KNN and PCA speedups reach 160x and 21x, respectively. The huge level of data dependencies in PCA makes it communication-bounded, while, in the case of SVM, there is no data dependencies, so the speedup is almost the ideal one when using 256 cores in parallel.

It should be noted that, to obtain the global figures (*Time* and *Av. Power*), not only the time of the three algorithms is considered, but also the data management between stages and both initializing and shutting down the platform. Consequently, the power consumption of the sequential stages needs to be considered. That is the rationale behind the fact that, in the case of using small images, power metrics for the experiments using 16 PEs per cluster have an average power consumption lower than their 1 PE counterparts.

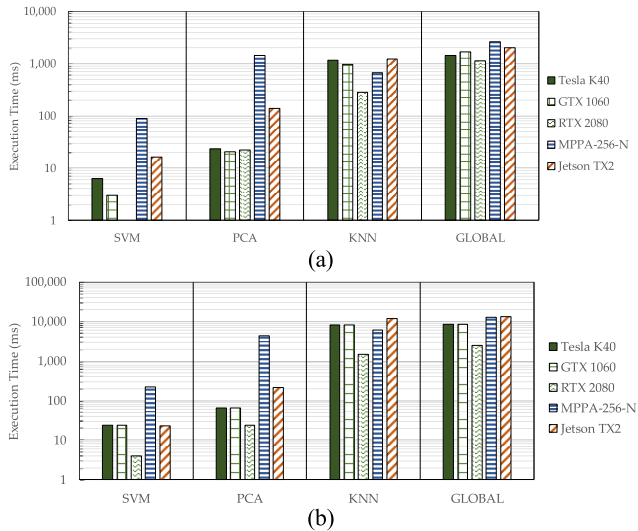

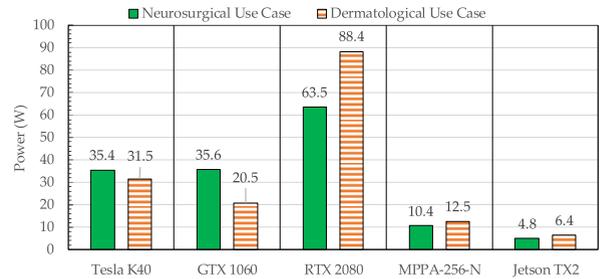

**FIGURE 5.** Average execution time comparison between the different platforms and configurations in the (a) neurosurgical use case and (b) dermatological use case. Representation in logarithmic scale.

**FIGURE 6.** Average power consumption comparison between the different platforms and configurations in both use cases.

Finally, regarding the FoMs, please note that the configuration in which all the MPPA-256-N resources are employed is always the best one regardless of the image that is employed.

### D. OVERALL RESULTS AND COMPARISON

Once the individual results have been presented, this section deals with the comparison among them. As mentioned before, this comparison is twofold: first, all the implementations are going to be compared in terms of performance; afterwards, they will be compared in terms of power consumption.

#### 1) PERFORMANCE COMPARISON

Fig. 5.a and b present the performance comparison for the neurosurgical and the dermatological use cases, respectively. For each case, three different images have been tested. To simplify the figures, their execution times have been averaged. Both figures have been divided in four slots: the first three provide the comparison of the execution times per algorithm; likely, the last one provides the comparison of the global times. Two important aspects should be highlighted:

- For the cases in which several versions have been described –i.e., MPPA-256-N and Jetson TX2–, the fastest versions have been selected for this comparison: ''16-16'' for the MPPA-256-N and ''MaxCLK'' for the Jetson TX2.
- To compare the results, a logarithmic scale has been used. Fig. 5 presents the execution times in milliseconds.

Generally speaking, it can be observed that, as expected, RTX 2080 is the platform providing the fastest results, since it is also the one with more resources. Likely, the MPPA-256-N provides the slowest results, which is also consistent, since the difference in the volume of resources among this platform and the rest under study is considerable. However, there are some interesting results that are worth commenting.

Regarding MPPA-256-N, even though in the global execution time it is the one with the slowest results, it is worth highlighting the extremes. On the one hand, for PCA algorithm, it can be observed that the results obtained are much worse than the rest. On the other hand, for KNN algorithm, it can be seen that, surprisingly, the results obtained with MPPA-256-N are beaten only by those obtained with RTX 2080. These results are the best example to characterize this platform: in the case of PCA, due to the memory limit of 2 MB per cluster, more than half of the executing time is spent on communicating data and synchronizing the clusters, while the cores are idle for long periods of time. This does not happen with the rest of the platforms: since they present bigger memories, their time spent on communications is considerably less, which has a direct impact on the execution time. However, in the case of KNN, the cores perform a huge number of operations requiring only one communication, which increases exponentially the platform performance. As a result, it can be concluded that MPPA can be really useful for algorithms requiring a considerable amount of computing operations, while maintaining the memory usage within a reasonable limit.

Finally, in general lines, another aspect that draws attention is the difference in the execution times between the desktop GPUs and the other platforms. This difference is remarkable for SVM and PCA, which is approximately one order of magnitude, while for KNN this difference does not exist anymore. This happens because PCA and SVM generically perform simple linear algebra operations which highly depend on the platform resources, while KNN involves more complex operations –e.g., sorting–. As a result, KNN features a really low instruction level parallelism, causing the threads to remain in an idle stay frequently and, hence, heavily degrading the GPU pipeline efficiency. It should be noted that this does not happen with the MPPA-256-N, since there is no pipeline in this platform.

#### 2) POWER CONSUMPTION COMPARISON

After comparing all the implementations and platforms in terms of performance, now they are compared in terms of power consumption. As with the previous comparison, for this one also the fastest versions of both MPPA-256-N and Jetson have been selected –i.e., ''16-16'' and ''MaxCLK''–. Also as happened before, the results for each set of images

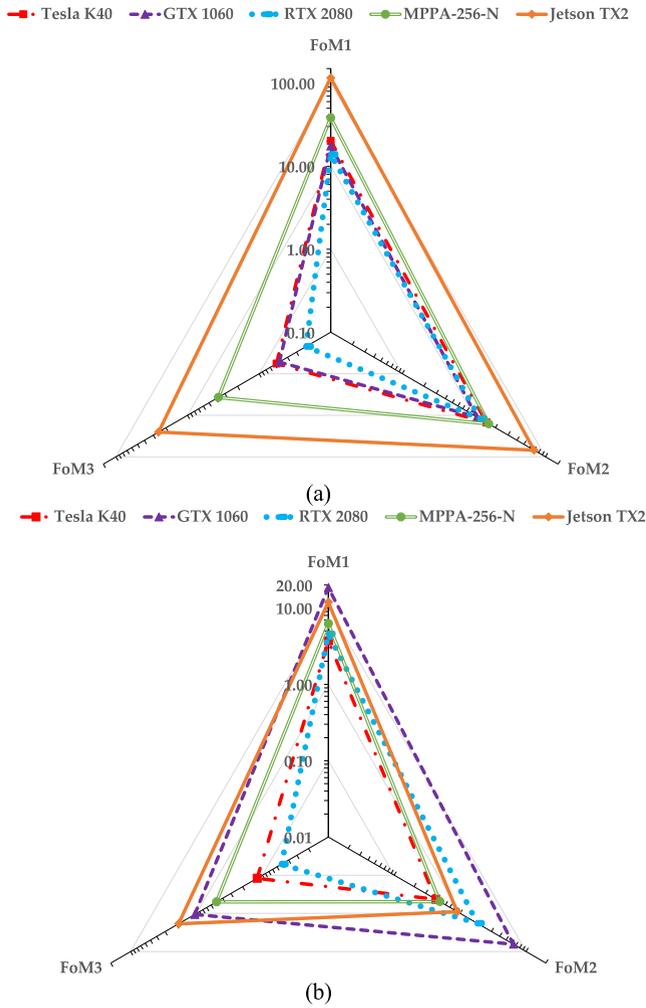

**FIGURE 7.** Average FoMs comparison between platforms and configurations (representation in logarithmic scale) for the two use cases. (a) Neurosurgical use case results. (b) Dermatological use case results with the logarithmic scale limited to 20 for legibility purposes (the higher the FoM, the best performance the implementation offers).

have been averaged to present one result per use case, as shown in Fig. 6.

On the one hand, as expected the three desktop GPUs present higher power consumption than both MPPA-256-N and Jetson TX2, especially RTX 2080, which is also the one with more resources usage. On the other hand, regarding MPPA-256-N and Jetson TX2, both of them were chosen precisely because of their energy efficiency, which is shown in this study. Both of them present results far from the ones obtained with the desktop GPUs, which makes them suitable for accelerating applications with low power consumption requirements. Specifically, among the platforms selected in this research work, Jetson TX2 is the one with the best balance between energy efficiency and performance, since it achieves fast results while maintaining the power consumption within reasonable limits.

### 3) GLOBAL COMPARISON

To conclude this discussion, Fig. 7 provides a global comparison, for both use cases, of the FoMs described at the beginning of this section. As a reminder, all of them link processing time and power consumption: $FoM\,1$ is a general index, $FoM\,2$ gives more importance to the time and $FoM\,3$ favors the power consumption. In any case, the higher the FoM, the better the performance the implementation offers. As it can be observed, in general lines the best FoMs are obtained for Jetson TX2, which is consistent with the conclusion obtained from the previous discussion.

Furthermore, we can also see that the size of the images has an important impact on the performance of the platform. On the one hand, for the neurosurgical use case –i.e., the one with smaller images– both MPPA-256-N and Jetson TX2 present better $FoMs$ than the rest (Fig. 7.a), but it turns around for the dermatological use case (Fig. 7.b). This is easily explained with the memory resources of the GPUs: the larger the images to process, the higher the performance. However, for the low-power platforms, the resources are already fully exploited and more iterations are required; thus, no performance improvement is achieved.

On the other hand, for the dermatological use case –i.e., the one with larger images–, it can be observed that, depending on the $FoM$, GTX 1060 starts to outperform Jetson TX2, becoming also a good option.

## V. CONCLUSION

This paper presents an evaluation of the performance and power consumption of two different HS medical applications implemented onto different HPC architectures. These applications have been chosen because they have different constraints: the neurosurgical use case requires real-time processing, since its objective is to help surgeons detect brain tumors during surgical procedures, while the dermatological use case favors low-power consumption, as one of its main features is to be an independent portable system.

To evaluate the applications, three different HS images have been chosen for each use case (available in the supplementary material), and they have been processed using a SS classification approach, composed of three different algorithms: PCA for dimensionality reduction, SVM to provide a classification, and KNN to spatially filter the results.

The obtained results show that, as initially expected, the three desktop GPUs are the ones providing fastest results, since they are the ones providing more computational resources; however, their energy consumption makes them not suitable to be used in applications where low-power consumption is compulsory.

In that case, both MPPA-256-N and Jetson TX2 present themselves as alternatives. On the one hand, MPPA-256-N is a manycore platform specifically thought for low-power consumption applications. As observed in the results, although in general terms it cannot compete with GPUs performance capabilities, this platform provides surprisingly good results for applications requiring substantial amounts of operations while keeping the memory usage within reasonable limits, since it is its main limitation.

On the other hand, Jetson TX2 comes from a family of GPUs with less computational resources, but also with considerably less power consumption. As shown in Section IV, the great difference of power consumption between the desktop GPUs and the Jetson TX2 clearly outgrows the difference in performance, which makes this platform the most suitable for both applications. However, when there are no power limitations and very tight real-time requirements, the optimal platform is the RTX 2080 GPU.

## ACKNOWLEDGMENT

The authors would like to thank NVIDIA Corporation for the donation of the NVIDIA Tesla K40 GPU used for this research.

## REFERENCES

[1] R. Smith, "Introduction to hyperspectral imaging," *Microimages. Retrieved*, Jun. 2006, pp. 1–24.

[2] M. Govender, K. Chetty, and H. Bulcock, "A review of hyperspectral remote sensing and its application in vegetation and water resource studies," *Water SA*, vol. 33, no. 2, pp. 145–152, 2007.

[3] G. J. Edelman, E. Gaston, T. G. van Leeuwen, P. J. Cullen, and M. C. G. Aalders, "Hyperspectral imaging for non-contact analysis of forensic traces," *Forensic Sci. Int.*, vol. 223, nos. 1–3, pp. 28–39, 2012.

[4] G. Elmasry, M. Kamruzzaman, D. W. Sun, and P. Allen, "Principles and applications of hyperspectral imaging in quality evaluation of agro-food products: A review," *Crit. Rev. Food Sci. Nutrition*, vol. 52, no. 11, pp. 999–1023, Jul. 2012.

[5] G. Bonifazi, R. Palmieri, and S. Serranti, "Hyperspectral imaging applied to end-of-life (EOL) concrete recycling," *Technisches Messen*, vol. 82, no. 12, pp. 616–624, Dec. 2015.

[6] M. A. Calin, S. V. Parasca, D. Savastru, and D. Manea, "Hyperspectral imaging in the medical field: Present and future," *Appl. Spectrosc. Rev.*, vol. 49, no. 6, pp. 435–447, 2014.

[7] G. Lu and B. Fei, "Medical hyperspectral imaging: A review," *Proc. SPIE*, vol. 19, no. 1, Jan. 2014, Art. no. 10901.

[8] S. Ortega, H. Fabelo, D. K. Iakovidis, A. Koulaouzidis, and G. M. Callico, "Use of hyperspectral/multispectral imaging in gastroenterology. Shedding some–different–light into the dark," *J. Clin. Med.*, vol. 8, no. 1, p. 36, Jan. 2019.

[9] M. Halicek, J. V. Little, X. Wang, A. Y. Chen, and B. Fei, "Optical biopsy of head and neck cancer using hyperspectral imaging and convolutional neural networks," *J. Biomed. Opt.*, vol. 24, no. 3, p. 1, Mar. 2019, Art. no. 036007.

[10] S. Ortega, H. Fabelo, R. Camacho, M. de la Luz Plaza, G. M. Callicó, and R. Sarmiento, "Detecting brain tumor in pathological slides using hyperspectral imaging," *Biomed. Opt. Express*, vol. 9, no. 2, pp. 818–831, Feb. 2018.

[11] E. L. P. Larsen, L. L. Randeberg, E. Olstad, O. A. Haugen, A. Aksnes, and L. O. Svaasand, "Hyperspectral imaging of atherosclerotic plaques *in vitro*," *Proc. SPIE*, vol. 16, no. 2, Feb. 2011, Art. no. 026011.

[12] D. R. McCormack, A. J. Walsh, W. Sit, C. L. Arteaga, J. Chen, R. S. Cook, and M. C. Skala, "*In vivo* hyperspectral imaging of microvessel response to trastuzumab treatment in breast cancer xenografts," *Biomed. Opt. Express*, vol. 5, no. 7, pp. 2247–2261, 2014.

[13] H. Fabelo, "*In-vivo* hyperspectral human brain image database for brain cancer detection," *IEEE Access*, vol. 7, pp. 39098–39116, 2019.

[14] J. M. Benavides, S. Chang, S. Y. Park, R. Richards-Kortum, N. Mackinnon, C. MacAulay, A. Milbourne, A. Malpica, and M. Follen, "Multispectral digital colposcopy for *in vivo* detection of cervical cancer," *Opt. Express*, vol. 11, no. 10, pp. 1223–1236, 2003.

[15] H. Fabelo, M. Halicek, S. Ortega, M. Shahedi, A. Szolna, J. F. Piñeiro, C. Sosa, A. J. O'Shanahan, S. Bisshopp, C. Espino, M. Márquez, M. Hernández, D. Carrera, J. Morera, G. M. Callico, R. Sarmiento, and B. Fei, "Deep learning-based framework for *in vivo* identification of glioblastoma tumor using hyperspectral images of human brain," *Sensors*, vol. 19, no. 4, p. 920, Feb. 2019.

[16] G. Lu, L. Halig, D. Wang, Z. G. Chen, and B. Fei, "Hyperspectral imaging for cancer surgical margin delineation: Registration of hyperspectral and histological images," *Proc. SPIE, Med. Imag. 2014, Image-Guided Procedures, Robot. Intervent., Model.*, vol. 9036, Mar. 2014, Art. no. 90360T.

[17] B. Regeling, B. Thies, A. O. H. Gerstner, S. Westermann, N. A. Müller, J. Bendix, and W. Laffers, "Hyperspectral imaging using flexible endoscopy for laryngeal cancer detection," *Sensors*, vol. 16, no. 8, p. 1288, Aug. 2016.

[18] H. Fabelo et al., "Spatio-spectral classification of hyperspectral images for brain cancer detection during surgical operations," *PLoS ONE*, vol. 13, no. 3, Mar. 2018, Art. no. e0193721.

[19] H. Fabelo et al., "An intraoperative visualization system using hyperspectral imaging to aid in brain tumor delineation," *Sensors*, vol. 18, no. 2, p. 430, 2018.

[20] M. Halicek, H. Fabelo, S. Ortega, G. M. Callico, and B. Fei, "*In-vivo* and ex-vivo tissue analysis through hyperspectral imaging techniques: Revealing the invisible features of cancer," *Cancers*, vol. 11, no. 6, p. 756, May 2019.

[21] M. Li, S. Zang, B. Zhang, S. Li, and C. Wu, "A review of remote sensing image classification techniques: The role of Spatio-contextual information," *Eur. J. Remote Sens.*, vol. 47, no. 1, pp. 389–411, 2014.

[22] D. Ravi, H. Fabelo, G. M. Callico, and G. Yang, "Manifold embedding and semantic segmentation for intraoperative guidance with hyperspectral brain imaging," *IEEE Trans. Med. Imag.*, vol. 36, no. 9, pp. 1845–1857, Sep. 2017.

[23] G. Camps-Valls and L. Bruzzone, "Kernel-based methods for hyperspectral image classification," *IEEE Trans. Geosci. Remote Sens.*, vol. 43, no. 6, pp. 1351–1362, Jun. 2004.

[24] C. H. Lee and H.-J. Yoon, "Medical big data: Promise and challenges," *Kidney Res. Clin. Pract.*, vol. 36, no. 1, pp. 3–11, Mar. 2017.

[25] K. Huang, S. Li, X. Kang, and L. Fang, "Spectral–spatial hyperspectral image classification based on KNN," *Sens. Imag.*, vol. 17, no. 1, pp. 1–13, Dec. 2016.

[26] V. Zheludev, I. Pölönen, N. Neittaanmäki-Perttu, A. Averbuch, P. Neittaanmäki, M. Grönroos, and H. Saari, "Delineation of malignant skin tumors by hyperspectral imaging using diffusion maps dimensionality reduction," *Biomed. Signal Process. Control*, vol. 16, pp. 48–60, Feb. 2015.

[27] N. Chiang, J. K. Jain, J. Sleigh, and T. Vasudevan, "Evaluation of hyperspectral imaging technology in patients with peripheral vascular disease," *J. Vascular Surg.*, vol. 66, no. 4, pp. 1192–1201, Oct. 2017.

[28] S. Saidi, R. Ernst, S. Uhrig, H. Theiling, and B. D. De Dinechin, "The shift to multicores in real-time and safety-critical systems," in *Proc. Int. Conf. Hardw./Softw. Codesign Syst. Synth. (CODES+ISSS)*, Oct. 2015, pp. 220–229.

[29] F. Melgani and L. Bruzzone, "Classification of hyperspectral remote sensing images with support vector machines," *IEEE Trans. Geosci. Remote Sens.*, vol. 42, no. 8, pp. 1778–1790, Aug. 2004.

[30] T.-F. Wu, C.-J. Lin, and R. C. Weng, "Probability estimates for multi-class classification by pairwise coupling," *J. Mach. Learn. Res.*, vol. 5, pp. 975–1005, Aug. 2004.

[31] D. Madroñal, R. Lazcano, R. Salvador, H. Fabelo, S. Ortega, G. M. Callico, E. Juarez, and C. Sanz, "SVM-based real-time hyperspectral image classifier on a manycore architecture," *J. Syst. Archit.*, vol. 80, pp. 30–40, Oct. 2017.

[32] R. Lazcano, D. Madroñal, H. Fabelo, S. Ortega, R. Salvador, G. M. Callico, E. Juarez, and C. Sanz, "Adaptation of an iterative PCA to a manycore architecture for hyperspectral image processing," *J. Signal Process. Syst.*, pp. 1–13, May 2018.

[33] L. Ma, M. M. Crawford, X. Yang, and Y. Guo, "Local-manifold-learning-based graph construction for semisupervised hyperspectral image classification," *IEEE Trans. Geosci. Remote Sens.*, vol. 53, no. 5, pp. 2832–2844, May 2015.

[34] G. Florimbi, H. Fabelo, E. Torti, R. Lazcano, D. Madroñal, S. Ortega, R. Salvador, F. Leporati, G. Danese, A. Báez-Quevedo, G. M. Callicó, E. Juárez, C. Sanz, and R. Sarmiento, "Accelerating the K-nearest neighbors filtering algorithm to optimize the real-time classification of human brain tumor in hyperspectral images," *Sensors*, vol. 18, no. 7, p. 2314, Jul. 2018.

[35] NVIDIA. *cuBLAS Library Documentation, CUDA Toolkit Documentation*. Accessed: Jul. 12, 2019. [Online]. Available: https://docs.nvidia.com/cuda/cublas/index.html

[36] S. Rennich. *CUDA C/C++ Streams and Concurrency, NVIDIA Developer*. Accessed: Jul. 12, 2019. [Online]. Available: https://developer.download.nvidia.com/CUDA/training/StreamsAndConcurrencyWebinar.pdf

[37] NVIDIA. *cuSOLVER Lbrary Documentation, CUDA Toolkit Documentation*. Accessed: Jul. 12, 2019. [Online]. Available: https://docs.nvidia.com/cuda/cusolver/index.html

[38] V. Volkov, *Understanding Latency Hiding on GPUs*. Berkeley, CA, USA: UC Berkeley, 2016.

[39] M. Harris. (2007). *Optimizing Parallel Reduction in CUDA, Nvidia Developer Technology*. Accessed: Jun. 29, 2019. [Online]. Available: http://developer.download.nvidia.com/compute/cuda/1.1-Beta/x86_website/projects/reduction/doc/reduction.pdf

[40] T. Wang, L. Guo, G. Li, J. Li, R. Wang, M. Ren, and J. He, "Implementing the jacobi algorithm for solving eigenvalues of symmetric matrices with CUDA," in *Proc. 7th Int. Conf. Netw., Archit., Storage*, Jun. 2012, pp. 69–78.

[41] J. Hascoët, B. D. de Dinechin, P. G. de Massas, and M. Q. Ho, "Asynchronous one-sided communications and synchronizations for a clustered manycore processor," in *Proc. 15th IEEE/ACM Symp. Embedded Syst. Real-Time Multimedia*, Oct. 2017, pp. 51–60.

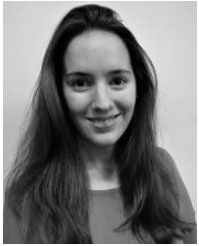

**RAQUEL LAZCANO** received the B.Sc. degree in communication electronics engineering and the M.Sc. degree in systems and services engineering for the information society from the Universidad Politécnica de Madrid (UPM), Spain, in 2014 and 2015, respectively, where she is currently pursuing the Ph.D. degree in systems and services engineering for the information society with the Electronic and Microelectronic Design Group (GDEM). In 2015, she stayed four months at the Institute of Electronics and Telecommunications of Rennes (IETR), National Institute of Applied Sciences (INSA), France, as an Interchange Student of the M.Sc. degree. She is the author or a coauthor of nine indexed journals and 17 contributions to technical conferences. Her research interests include high-performance multicore processing systems, real-time hyperspectral image processing, and the automatic optimization of the parallelism in real-time systems.

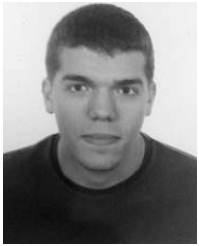

**DANIEL MADROÑAL** received the B.Sc. degree in communication electronics engineering and the M.Sc. degree in systems and services engineering for the information society from the Universidad Politécnica de Madrid (UPM), Spain, in 2014 and 2015, respectively, where he is currently pursuing the Ph.D. degree in systems and services engineering for the information society with the Electronic and Microelectronic Design Group (GDEM). In 2015, he stayed four months at the National Institute of Applied Sciences (INSA), France, as an Interchange Student of the M.Sc. degree. He is the author or a coauthor of nine indexed journals and 18 contributions to technical conferences. His research interests include high-performance multi- and many-core processing systems, real-time hyperspectral image processing, and the automatic optimization of the energy consumption in high-performance systems.

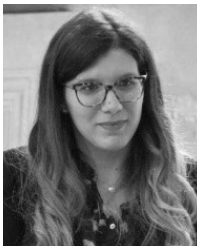

**GIORDANA FLORIMBI** was born in Teramo, Italy, in 1989. She received the bachelor's degree in biomedical engineering from the Università Politecnica delle Marche, Ancona, Italy, in 2012, and the master's degree in bioengineering and the Ph.D. degree in bioengineering and bioinformatics from the University of Pavia, Pavia, Italy, in 2015 and 2019, respectively, where she is currently a Postdoctoral Researcher with the Engineering Faculty. Her research interest includes real-time elaborations in the support medical systems development and in neuroscience, exploiting HPC technologies.

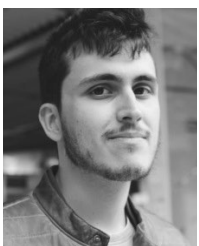

**JAIME SANCHO** received the B.Sc. degree in telecommunication engineering and the M.Sc. degree in systems and services engineering for the information society from the Universidad Politécnica de Madrid (UPM), Spain, in 2017 and 2018, respectively, where he is currently pursuing the Ph.D. degree in systems and services engineering for the information society with the Electronic and Microelectronic Design Group (GDEM). His research interests include high-performance graphics processing systems, real-time hyperspectral image processing, and immersive computer vision applications.

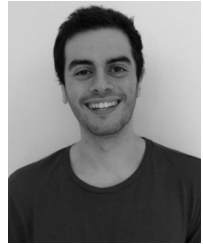

**SERGIO SANCHEZ** received the B.Sc. degree in sound and image engineering from the Universidad Politécnica de Madrid (UPM), Spain, in 2019, where he is currently a Researcher with the Electronic and Microelectronic Design Group (GDEM). His research interests include the design of hyperspectral image processing techniques and machine learning algorithms for real-time applications.

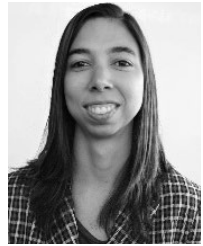

**RAQUEL LEON** received the Telecommunication Engineer degree and the research master's degree in telecommunication technologies from the University of Las Palmas de Gran Canaria, Spain, in 2017, and 2018, respectively. Since then, she has been conducting her research activity in the Integrated System Design Division, Institute for Applied Microelectronics, University of Las Palmas de Gran Canaria, in the field of electronic and bioengineering. In 2018, she started to work as a Researcher in the ITHaCA project (hyperspectral identification of brain tumors). Her current research interest includes the use of hyperspectral imaging for real-time cancer detection.

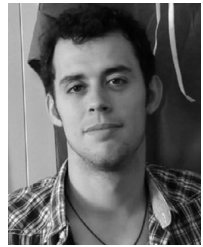

**HIMAR FABELO** received the Telecommunication Engineer degree and the Ph.D. degree in telecommunication technologies from the University of Las Palmas de Gran Canaria, Las Palmas de Gran Canaria, Spain, in 2014, and 2019, respectively. Since then, he has been conducting his research activity in the Integrated System Design Division, Institute for Applied Microelectronics, University of Las Palmas de Gran Canaria, in the field of electronic and bioengineering. In 2015, he started to work as a Coordination Assistant and a Researcher in the HELICoiD European project, co-funded by the European Commission. In 2018, he performed a research stay at the Department of Bioengineering, Erik Jonsson School of Engineering and Computer Science, The University of Texas at Dallas, collaborating with Prof. B. Fei in the use of medical hyperspectral imaging analysis using deep learning. His research interests include the use of machine learning and deep learning techniques applied to hyperspectral images to discriminate between healthy and tumor samples for human brain tissues in real-time during neurosurgical operations.

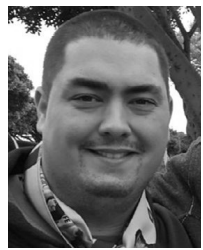

**SAMUEL ORTEGA** received the Telecommunication Engineer degree and the research master degree in telecommunication technologies from the University of Las Palmas de Gran Canaria, Spain, in 2014, and 2015, respectively. Since then, he has been conducting his research activity in the Integrated System Design Division, Institute for Applied Microelectronics, University of Las Palmas de Gran Canaria, in the field of electronic and bioengineering. In 2015, he started to work as a Coordination Assistant and a Researcher in the HELICoiD European project, co-funded by the European Commission. His current research interest includes the use of machine learning algorithms in medical applications using hyperspectral images.

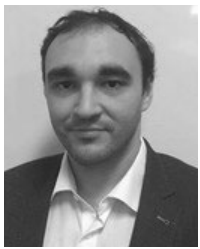

**EMANUELE TORTI** (M'13) was born in Voghera, Italy, in 1987. He received the bachelor's degree in electronic engineering, the master's degree *(cum laude)* in computer science engineering, and the Ph.D. degree in electronics and computer science engineering from the University of Pavia, Pavia, Italy, in 2009, 2011, and 2014, respectively, where he is currently an Assistant Professor with the Engineering Faculty. His research interest includes high-performance architectures for real-time image processing and signal elaboration.

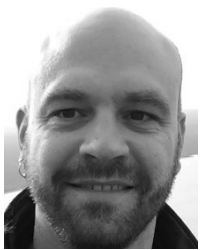

**RUBEN SALVADOR** received the Ph.D. degree in electrical and computer engineering from the Universidad Politécnica de Madrid (UPM), where he is currently an Assistant Professor with the Department of Telematics and Electronics Engineering and a Researcher with the Center on Software Technologies and Multimedia Systems for Sustainability (CITSEM-UPM). In 2009, he was a Visiting Research Student (for four months) with the Department of Computer Systems, Brno University of Technology. In 2017, he was a Visiting Professor (for five months) with IETR/INSA, Rennes. He was a Research Assistant with the Center of Industrial Electronics (CEI-UPM), from 2006 to 2011, and the Intelligent Vehicle Systems Division, University Institute for Automobile research (INSIA-UPM), from 2005 to 2006. He is the author/coauthor of around 40 peer-reviewed publications in international journals/conferences and one book chapter. His research interest includes high-performance and self-adaptive computer systems, with a particular focus in the design of reconfigurable and parallel heterogeneous accelerators for embedded systems. Applications of his work have included evolvable hardware for systems self-adaptation in harsh environments and acceleration of machine learning applied to hyperspectral image processing for cancer detection. He serves as a TPC Member for various international conferences and acts as a Reviewer for a number of international journals/conferences. He has participated in nine EU/national research projects and nine industrial projects.

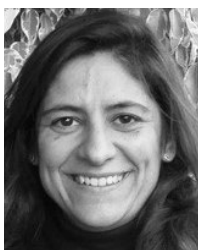

**MARGARITA MARRERO-MARTIN** received the M.S. degree in telecommunication engineering and the Ph.D. degree from the University of Las Palmas de Gran Canaria (ULPGC), Spain, in 2001 and 2012, respectively, where she is currently an Associate Professor. She researches with the Institute for Applied Microelectronics, Microelectronics Technology Division. Until 2017, her research lines of interest were the characterization, modeling, and design of RF integrated passive components. She has participated as a Researcher in national and European funded projects, coauthoring more than 50 articles in journal articles and conferences. In 2017, she changed her field of research to the area of bioengineering, specifically the use of hyperspectral imaging to detect cancer in real time. She has occupied different management positions at the ULPGC

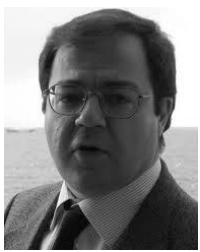

**FRANCESCO LEPORATI** received the Ph.D. degree in electronics and computer engineering from the University of Pavia, Pavia, Italy, in 1993, where he is currently an Associate Professor with the Industrial Informatics and Embedded Systems and Digital Systems Design, Engineering Faculty. His research interests include automotive applications, FPGA and application-specific processors, embedded real-time systems, and computational physics. He is also a member of the Euromicro Society and an Associate Editor of *Microprocessors and Microsystems*.

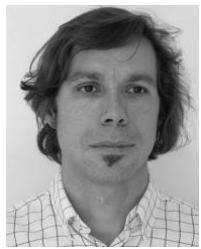

**EDUARDO JUAREZ** (M'96) received the Ph.D. degree from the EPFL, in 2003. From 1994 to 1997, he was a Researcher with the Digital Architecture Group, UPM, and a Visiting Researcher with ENST, Brest, France, and the University of Pennsylvania, Philadelphia, PA, USA. From 1998 to 2000, he was an Assistant with the Integrated Systems Laboratory (LSI), EPFL. From 2000 to 2003, he was a Senior Systems Engineer with the Design Centre, Transwitch Corp., Switzerland. In December 2004, he joined the GDEM as a Postdoctoral Researcher. Since 2007, he has been an Assistant Professor with UPM. He is a coauthor of one book and the author or a coauthor of more than 50 articles and contributions to technical conferences. His research interest includes solving, from a holistic perspective, the power/energy consumption optimization problem of multimedia handheld devices. He has participated in nine competitive research projects and 18 noncompetitive industrial projects.

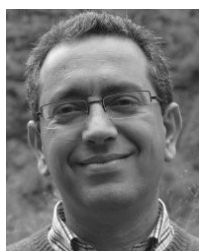

**GUSTAVO M. CALLICO** (M'08) received the M.S. degree (Hons.) in telecommunication engineering and the Ph.D. and the European Doctorate degrees (Hons.) from the University of Las Palmas de Gran Canaria (ULPGC), in 1995 and 2003, respectively, where he is currently an Associate Professor.

From 1996 to 1997, he was granted with a research grant from the National Educational Ministry. In 1997, he was hired by the university as an Electronic Lecturer. In 1994, he joined the Institute for Applied Microelectronics (IUMA). From 2000 to 2001, he stayed at the Philips Research Laboratories (NatLab), Eindhoven, The Netherlands, as a Visiting Scientist, where he developed his Ph.D. thesis. He currently develops his research activities in the Integrated Systems Design Division, Institute for Applied Microelectronics (IUMA). He has more than 110 publications in national and international journals, conferences, and book chapters. He has participated in 18 research projects funded by the European Community, the Spanish Government, and international private industries. Since 2015, he has been the responsible for the scientific-technological equipment project called Hyperspectral Image Acquisition System of High Spatial and Spectral Definition, granted by the General Directorate of Research and Management of the National Research and Development Plan, funded through the General Directorate of Scientific Infrastructure. He has been a Coordinator of the European project HELICoiD [Future and Emerging Technologies (FET)] under the Seventh Framework Program. His current research interests include hyperspectral imaging for real-time cancer detection, real-time super-resolution algorithms, synthesis-based design for SOCs and circuits for multimedia processing, and video coding standards, especially for H.264 and SVC. He has been an Associate Editor of the IEEE TRANSACTIONS ON CONSUMER ELECTRONICS, since 2009. He is also a Senior Associated Editor of the IEEE TRANSACTIONS ON CONSUMER ELECTRONICS. He has been an Associate Editor of IEEE ACCESS, since 2016.

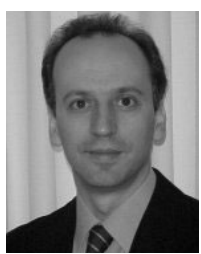

**CESAR SANZ** (S'87–M'88–SM'13) received the Ph.D. degree from the Universidad Politécnica de Madrid (UPM), Madrid, Spain, in 1998, where he is currently a Full Professor with the ETSIS de Telecomunicación. He has been the Director of the ETSIS de Telecomunicación for eight years. He also leads the Electronic and Microelectronic Design Group, UPM, where he has been involved in research and development Projects. Since 2013, he has been a Researcher with CITSEM. His current research interest includes microelectronic design applied to real-time image processing.

● ● ●